\documentclass[10pt]{article}
\usepackage{epsf}

\newlength{\dinwidth}
\newlength{\dinmargin}
\def\lapproxeq{\lower .7ex\hbox{$\;\stackrel{\textstyle <}{\sim}\;$}}
\def\gapproxeq{\lower .7ex\hbox{$\;\stackrel{\textstyle >}{\sim}\;$}}

\def\be{\begin{equation}}
\def\ee{\end{equation}}
\def\bea{\begin{eqnarray}}
\def\eea{\end{eqnarray}}

\def\GeV{{\rm GeV}}

\def\bb{{b\bar{b}}}

\begin{document}
\titlepage

\begin{flushright}
IPPP/04/13 \\
DCPT/04/26\\
15th March 2004 \\
\end{flushright}

\vspace*{4cm}

\begin{center}
{\Large \bf Double-diffractive $\chi$ meson production at the hadron colliders}

\vspace*{1cm} \textsc{V.A.~Khoze$^{a,b}$, A.D. Martin$^a$, M.G. Ryskin$^{a,b}$ and W.J. Stirling$^{a,c}$} \\

\vspace*{0.5cm} $^a$ Department of Physics and Institute for
Particle Physics Phenomenology, \\
University of Durham, DH1 3LE, UK \\[0.5ex]
$^b$ Petersburg Nuclear Physics Institute, Gatchina,
St.~Petersburg, 188300, Russia \\[0.5ex]
$^c$ Department of Mathematical Sciences, 
University of Durham, DH1 3LE, UK \\%
\end{center}

\vspace*{1cm}

\begin{abstract}
The double-diffractive production of $\chi_c$ and $\chi_b$ mesons, with
a rapidity gap on either side, is studied, using both the Regge
formalism and the perturbative QCD approach. Due to the rather low
scale, the exclusive double-diffractive 
process $pp\to p\; +\; \chi\; +\; p$ is predicted to dominate, whereas
the probability that the incoming protons dissociate is expected to be
relatively small.  We
evaluate the corresponding $\chi$ production cross sections at the Tevatron and LHC
energies. For the double-diffractive process with proton dissociation, it 
is possible to select events with large transverse
momenta transferred through the rapidity gaps, by measuring the transverse energy,
$E_\perp$, flows in the proton fragmentation regions. Then the large $E_\perp$
provides a scale to justify the use of perturbative QCD, and to allow
a spin-parity analysis of the centrally produced system to be performed, by studying the
 azimuthal angular correlations between the directions of
the forward and backward $E_\perp$ flows.  The central production of the 
new $X(3872)$ charmonium state is considered.  
\end{abstract}


\section{Introduction}

Central exclusive double-diffractive processes have traditionally been
considered as a promising way to study new (and old) particles in
an especially clean environment, see, for example, Ref.~\cite{DR}.
In addition they give information about the structure of the
Pomeron and of the mechanism of Pomeron-Pomeron fusion.  Here we
are particularly interested in the double-diffractive production of, C-even, heavy
quarkonium $(\chi_c,\chi_b)$ at the Tevatron and the
LHC \cite{KMRmm,KMRProsp}, see also Refs.~\cite{Pump,FY}.  Since
the early days of QCD, heavy quark production has been a fertile
testing ground of many aspects of the theory.  More recently,
there has been much activity in studying the long standing
discrepancy between the NLO predictions and the data for heavy
quark production at hadron colliders \cite{MC}.   Heavy quarkonium
production is a valuable tool, since it provides important
information on the physics of bound states and, in particular,
allows a test of the ideas and methods of QCD effective field
theories, which is one of the most popular recent approaches; for
reviews and references see, for example, \cite{EFT}.

Detailed experimental and theoretical studies of double-diffractive $\chi$ production
offer a new window on this topic, and may shed light on some of the
unresolved issues. These processes have some surprising features. 
In order to gain insight into the problems encountered in the evaluation of
these processes, we first recall an analogous process, which in many respects
is theoretically simpler.  Namely,
the exclusive double-diffractive central production of
Higgs bosons, which has been advocated as a good way to study the Higgs sector at the LHC \cite{DKMOR}.
Both the $\chi$ and Higgs production processes may be written in the form

\be pp\to p~+~(\chi~{\rm or}~ H)~+~p,
\label{eq:A1}
\ee
where the + signs are used to represent the presence of large rapidity gaps.
Since the Higgs is expected to have mass, $M_H$, about 120 GeV or more,
its production cross section
may be calculated perturbatively, via the diagram shown in Fig.~1.  The hard
subprocess $gg \to H$ is initiated by gluon-gluon fusion, and the second
$t$-channel gluon is needed to screen the colour.  The crucial observation is that
the integration over $Q_\perp$, going round the gluon loop, is made infrared safe by
Sudakov factors which ensure that the subprocess gluons remain untouched
in their evolution up to the hard scale $M_H/2$.  Provided forward proton
taggers are installed, the process has the advantage that the Higgs mass may
be measured in two independent ways: first from its decay products and second by observing
the forward protons which accurately determine the
`missing mass' $M_H$.  The predicted event rate and the signal-to-background ratio make
such a Higgs search feasible at the LHC, but unfortunately the rate is estimated
to be too small for this process to be of use at the Tevatron \cite{KMR,DKMOR}.

\begin{figure}
\begin{center}
\centerline{\epsfxsize=10cm\epsfbox{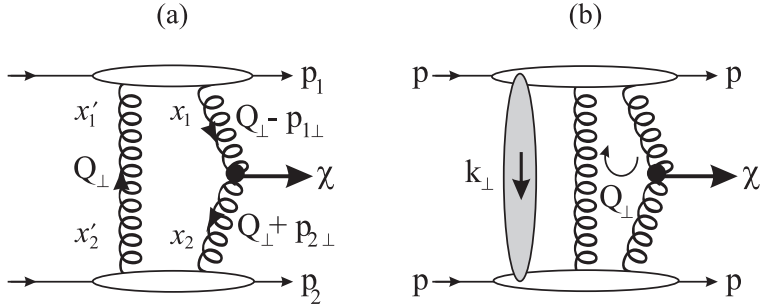}} \caption{(a)~The QCD diagram for double-diffractive exclusive
production of a $\chi (0^+)$ meson, $pp\to p + \chi + p$, where the gluons of the hard subprocess $gg\to \chi$ are colour
screened by the second $t$-channel gluon.~~(b)~The rescattering or absorptive corrections to $pp\to p + \chi + p$,
where the shaded region represents the soft $pp$ rescattering corrections, leading to the suppression factor
$\hat{S}^2$.\label{fig:3}}
\end{center}
\end{figure}

On the other hand the rate of double-diffractive $\chi$ production at the
Tevatron is expected to be much larger.  Indeed preliminary studies
indicate that such events may have already been detected \cite{CDFchi}.
Unfortunately such data cannot provide a reliable check on the perturbative
prediction of double-diffractive Higgs production.
Unlike the Higgs, the $\chi_c$, and to a lesser extent the $\chi_b$, are not sufficiently massive to make the
double-diffractive cross section predictions
infrared stable.  Moreover, the predictions depend on the gluon
distribution to the fourth power in the region $x \sim 10^{-4}$ and $Q^2 \sim 1~{\rm GeV}^2$,
where it is not sufficiently reliably known.  Nevertheless,
motivated by the forthcoming experimental data, and by special features
of the production process, it is informative to study double-diffractive $\chi_c$ and
$\chi_b$ production in more detail.  In addition, the interest in double-diffractive
$\chi_b$ (and even $\chi_c$) production has recently increased, since these processes may 
be measured at the LHC with luminosities of $10^{30} - 10^{31}~
{\rm cm}^{-2} {\rm sec}^{-1}$ by the CMS and TOTEM collaborations, with the special optics
of the TOTEM detector \cite{TOTEM}.
These forthcoming, and planned, measurements will sharpen our understanding of the physics
of these types of processes, as well as helping to find better ways to select
double-diffractive events.

Since $M_\chi$ is not large enough to justify the use of perturbative QCD, we will start
our discussion using a Regge framework.   We show how this formalism can be matched to
the perturbative approach, and we will use perturbative QCD to evaluate the
Pomeron-Pomeron $\to \chi$ fusion vertex.

Besides the pure exclusive process $pp\to p+\chi+p$, we will also
study central diffractive inclusive $\chi$
production in which both of the incoming protons are destroyed,
\be pp\to X+\chi+Y.
\label{eq:A2}
\ee
Recall that for a heavy centrally produced state, such as a Higgs boson, the cross
section for inclusive double-diffractive production is much larger than that for exclusive
double-diffractive production \cite{KMR,KMRProsp},
\be \sigma_{\rm incl} \gg \sigma_{\rm excl}.
\label{eq:A3}
\ee
The situation is quite different for the production of the relatively light $\chi$ states.
In this case we expect
exclusive production to dominate.   The reason is explained at the end of the next Section.
Another interesting feature is that we can justify the use of perturbative QCD for inclusive
double-diffractive $\chi$ production by selecting events with large momentum
transferred through the exchanged Pomerons (which can be measured as the
tranverse energy flows, $E_{iT}$, in the proton fragmentation regions).  These $E_T$'s provide
the large scale which clearly allows the use of the perturbative QCD formalism.

\section{The Regge framework for double-diffractive $\chi$ production}

Exclusive double-diffractive $\chi$ production by Pomeron-Pomeron fusion is
shown schematically in Fig.2(a), where $p_{1,2\perp}$ are the transverse momenta
of the outgoing protons,
\begin{equation}
 \frac{d\sigma}{d^2p_{1\perp}d^2p_{2\perp}dy}~~=~~\frac{1}{2^8\pi^5}
g_N^2(p^2_{1\perp})~V^2(p^2_{1\perp},p^2_{2\perp})~g_N^2(p^2_{2\perp})~
x_1^{2(1-\alpha_P(t_1))}x_2^{2(1-\alpha_P(t_2))},\label{eq:PPfusion}
\end{equation}
where $g_N$ is the Pomeron-nucleon coupling and $V$ is the Pomeron-Pomeron-$\chi$
fusion vertex.  The Pomeron trajectory is $\alpha_P(t)$, with $t_i\simeq -p^2_{i\perp}$, and
\begin{equation}
x_{1,2}~=~\frac{M_{\chi \perp}}{\sqrt s}~e^{\pm y}, \label{eq:x1x2}
\end{equation}
where $y$ is the c.m. rapidity of the produced $\chi$ meson, and
\begin{equation}
M^2_{\chi \perp}~=~M^2_\chi~+~({\vec p}_{1\perp}+{\vec p}_{2\perp})^2. \label{eq:M}
\end{equation}
For simplicity we have neglected the real part of the Pomeron amplitude,
assuming\footnote{In other words, we neglect the signature factors in (\ref{eq:PPfusion}) and (\ref{eq:PPP}),
assuming that $\alpha_P(t)$ is rather close to 1.} that Re/Im $\ll1$.
The coupling $g_N$ is known from  $pp$ data,
$$ \sigma^{\rm tot}(pp) ~=~g_N^2(0)~(s/m_N^2)^{a_P(0)-1},$$
 and $\alpha_P(t)$ can be taken either
from $J/\psi$ diffractive photoproduction data or from an analysis of $pp$ data.   The vertex $V$
will be discussed later.

\begin{figure}
\begin{center}
\centerline{\epsfxsize=10cm\epsfbox{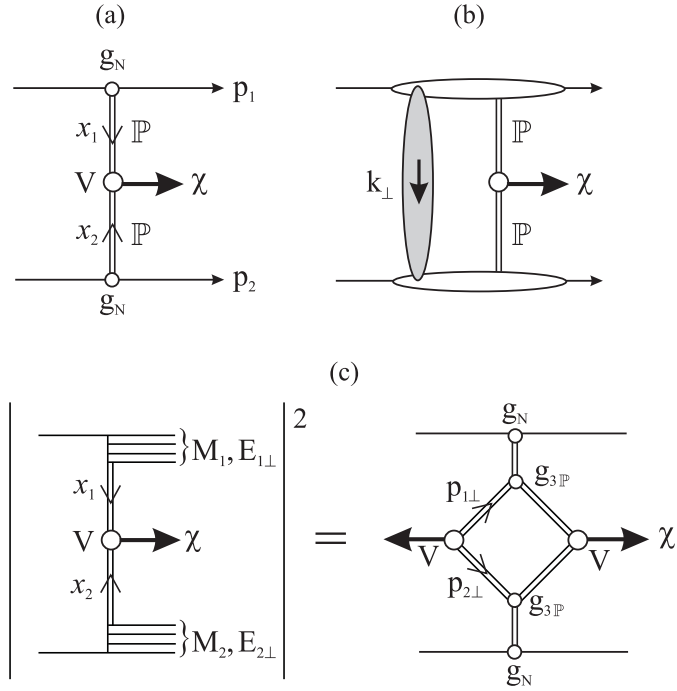}} \caption{(a)~The
Pomeron-Pomeron fusion diagram for double-diffractive exclusive
production of a $\chi (0^+)$ meson, $pp\to p + \chi + p$; see
(\ref{eq:PPfusion}).~~(b)~The rescattering or absorptive
corrections to diagram (a), where the shaded region represents the
soft $pp$ rescattering corrections, leading to the suppression
factor $\hat{S}^2$.~~(c) The amplitude (squared) for inclusive
double-diffractive $\chi$ production; the final diagram shows how
high-mass diffraction is given in terms of the triple-Pomeron
interaction; see (\ref{eq:PPP}).\label{fig:2}}
\end{center}
\end{figure}

For inclusive double-diffractive $\chi$ production (Fig.~2(c)) we treat low- and high-mass proton dissociation differently.
For low-mass dissociation, say $N \to N^*$, we just need
to replace the coupling $g_N(t)$ by $g_{N \to N^*}(t)$.
On the other hand the dissociation into a high-mass system is described in terms of the
triple-Pomeron interaction with coupling $g_{3P}$, see Fig.~2(c).   The corresponding cross section is
\begin{eqnarray}
 \frac{M^2_1 M^2_2 d\sigma}{dM^2_1dM^2_2d^2p_{1\perp}d^2p_{2\perp}dy}~~=~~\frac{1}{2^8\pi^7}
g_N^2(0)~ g_{3P}(p^2_{1\perp})~ V^2(p^2_{1\perp},p^2_{2\perp})~g_{3P}(p^2_{2\perp})~
 \left( \frac{M^2_1 M^2_2}{s_0^2}\right)^{\alpha_P(0)-1}
\nonumber \\
\left( \frac{x_2s}{M_1^2}\right)^{2(\alpha_P(p^2_{1\perp})-1)}
\left( \frac{x_1s}{M_2^2}\right)^{2(\alpha_P(p^2_{2\perp})-1)}.\label{eq:PPP}
\end{eqnarray}
Due to the condition
\begin{equation}
x_1x_2s~=~M^2_{\chi\perp}, \label{eq:cond}
\end{equation}
we see that the factor
\begin{equation}
\left( \frac{x_2s}{M_1^2}\right)^{2(\alpha_P(p^2_{1\perp})-1)}~~\propto~~x_1^{2(1-\alpha_P(p^2_{1\perp}))},
\label{eq:vv}
\end{equation}
and vice-versa.  Thus the cross sections of both the exclusive and inclusive diffractive dissociation
processes have the same energy and rapidity dependences, see (\ref{eq:x1x2}).  For fixed masses
$M_{1,2}$ (and fixed $p_{1\perp},p_{2\perp}$ and $y$) we have
\begin{equation}
d\sigma~~\propto~~s^{\alpha_P(p^2_{1\perp})+\alpha_P(p^2_{2\perp})-2}. \label{eq:sdep}
\end{equation}
However we have to include the absorptive corrections.  The gap survival factor
$\hat{S}^2$~---~that is the probability that the rapidity gaps will not be filled by
secondaries produced in the soft rescattering\footnote{See, for example, Ref.~\cite{KKMR},
and references therein.}~---~is shown schematically by the
graphs in Figs.~1(b) and 2(b).  The corresponding amplitudes interfere
destructively with the original bare amplitude.  The elastic amplitude, shown
by the shaded blob in Figs.~1(b) and 2(b), grows with energy and hence leads to a decrease
of $\hat{S}^2$.  The absorption is stronger for exclusive production (Fig.~1)
since it is harder to transfer the loop momentum $k_\perp$ through the
multiparticle chain formed by the diffractive excitations $M_1$ and $M_2$ \cite{KMRsoft}
\footnote{In impact parameter $b_\perp$ space, the elastic amplitude $A_{\rm el}(b_\perp)$
arising from the inelastic channels, via unitarity, is given by the square of the
inelastic matrix element, $A_{\rm el}(b_\perp)\propto |A_{\rm inel}(b_\perp)|^2,$
and is therefore concentrated at smaller $b_\perp$ where the absorption is stronger;
for more details see, for example, \cite{KMRtagg} and references therein.}.

First, we calculate the rapidity gap survival factors $\hat{S}^2$ for exclusive and inclusive
double diffractive $\chi_c$ production using the formalism of Ref.~\cite{KMRsoft}.
If we were to neglect the slope of the Pomeron trajectory, $\alpha_P^{'}=0$, then we find
\begin{equation}
\hat{S}^2({\rm Tevatron})~=~ 0.05~(0.15) \label{eq:S2Tev}
\label{eq:S1}
\end{equation}
\begin{equation}
\hat{S}^2 ({\rm LHC})~=~0.024~(0.10)\label{eq:S2LHC}
\end{equation}
for exclusive (inclusive) production.  When we include the Pomeron slope,
$\alpha_P^{'}=0.1~{\rm GeV}^2$ (which is consistent with the HERA
$J/\psi$ photoproduction data \cite{Jdata}),
the above survival factors become
\begin{equation}
\hat{S}^2({\rm Tevatron})~=~ 0.07~(0.16) \label{eq:S2aTev}
\end{equation}
\begin{equation}
\hat{S}^2 ({\rm LHC})~=~0.04~(0.11);\label{eq:S2aLHC}
\end{equation}
that is, in the Tevatron-LHC energy interval we may approximate the energy dependence by
\begin{equation}
\hat{S}^2 ~\propto~s^{-\delta}~~~{\rm with}~~~\delta~=~0.16~(0.1). \label{eq:S2sdep}
\end{equation}
For inclusive production we select events with rapidity gaps $\Delta \eta \geq 3$: so
$\alpha_P^{'}$ acts on a smaller rapidity interval, and therefore is not so
effective in enlarging the value of $\hat{S}^2$.

Recall that the triple-Pomeron vertex $g_{3P}(0)$ is small numerically in comparison with the elastic
Pomeron-nucleon coupling $g_N(0)$; see, for example, \cite{ABKK}.  
On the other hand $g_{3P}$ is almost independent of the values
of $p_{1\perp}$ and $p_{2\perp}$.   Thus the suppression $({g_{3P}/g_N})^2$ is partly compensated by the
larger interval of the transverse momenta $p_{i\perp}$ sampled in diffractive dissociation\footnote{For
the inclusive
process, the momentum transfers $p_{i\perp}$ can be determined by experimentally measuring the transverse energy
flows, $E_{iT}$, in the proton fragmentation regions.}.
To obtain a naive estimate of the ratio of the cross section of exclusive to inclusive $\chi $ production, we first note
that the ratio of cross section for single proton dissociation to that for elastic scattering
satisfies \cite{Kaid,KMRsoft}
\begin{equation}
R~=~\sigma_{\rm SD}/\sigma_{\rm el}~<~1/2.\label{eq:SDel}
\end{equation}
Thus for the `soft' process we expect
\begin{equation}
\sigma_{\rm incl}/ \sigma_{\rm excl}~\simeq~R^2~<~1/4.\label{eq:inexcl}
\end{equation}
As a consequence, even without tagging the forward protons, we will mainly observe the exclusive
$pp \to p~+~\chi~+~p$ process simply by selecting events with rapidity gaps\footnote{As
we will discuss below, this is not true for heavy boson production, where we have large Sudakov $T$-factor
suppression in order to ensure that the rapidity gaps survive against QCD radiation in the hard subprocess.
Since the Sudakov suppression is much stronger for the exclusive amplitude for, say, Higgs production,
we have $\sigma_{\rm incl} \gg \sigma_{\rm excl}$.}.

\section{The perturbative QCD approach to double-diffractive $\chi$ production}

To calculate the cross section for the central exclusive double-diffractive production of $\chi(0^+)$
states we use the formalism of Refs.~\cite{KMR,KMRmm,KMRProsp}.  The amplitudes are described by
the diagram shown in Fig.~\ref{fig:3}(a), where the hard subprocess $gg\to \chi(0^+)$ is initiated by
gluon--gluon fusion and where the second $t$-channel gluon is needed to screen the colour flow across the rapidity
gap intervals.
Ignoring, for the moment, the screening corrections of Fig.~\ref{fig:3}(b), the Born amplitude is of the
form~\cite{KKMRCentr}
\be T = A\pi^2\int\frac{d^2Q_\perp\ P(\chi (0^+))}{Q^2_\perp (\vec Q_\perp - \vec p_{1\perp})^2(\vec Q_\perp + \vec
p_{2\perp})^2}\: f_g(x_1, x_1', Q_1^2, \mu^2; t_1)f_g(x_2,x_2',Q_2^2,\mu^2; t_2), \label{eq:rat3} \ee
where the $gg\to \chi (0^+)$ subprocess is specified by \cite{KMRProsp}
\begin{equation}
 A^2=8\pi\Gamma(\chi \to gg)/M^3_{\chi}, \label{eq:Asquared}
\end{equation}
  and the polarisation factor\footnote{For $\chi (0^-)$ production the factor
$P({\chi (0^-)}) =  \left( (\vec Q_\perp - \vec p_{1\perp}) \times (\vec Q_\perp + \vec p_{2\perp})\right)\cdot \vec
n_0$,
where $\vec{n}_0$ is a unit vector in the beam direction; see, for example, \cite{KKMRCentr}.}
\be
 P({\chi (0^+)}) = (\vec Q_\perp - \vec p_{1\perp}) \cdot (\vec Q_\perp + \vec p_{2\perp}). \label{eq:rat4}
\ee
Here $\Gamma(\chi \to gg)$ is the width including the NLO corrections (that is the $K$ factor).
We assume the same NLO correction for the $gg \to \chi$ vertex as for the $\chi \to gg$
width, which can be valid only within a certain approximation.

The $f_g$'s in (\ref{eq:rat3}) are the skewed unintegrated gluon densities of the proton at the hard scale $\mu$, taken
typically to be
$M_{\chi}/2$, with
\be\begin{array}{l l l} Q_1 & = & \min\left\{Q_\perp,|(\vec Q_\perp - \vec p_{1\perp})|\right\},  \nonumber\\
Q_2 & = & \min\left\{Q_\perp,|(\vec Q_\perp + \vec p_{2\perp})|\right\}. \label{eq:rat5}
\end{array} \ee
The longitudinal momentum fractions carried by the gluons satisfy
\begin{equation}
\left(x^\prime\sim\frac{Q_\perp}{\sqrt{s}}\right) \ll \left(x\sim\frac{M_{\chi}}{\sqrt{s}}\right) \ll 1.
\label{eq:lmf_ineq}
\end{equation}
Below, we assume factorization of the unintegrated distributions,
\be f_g(x,x',Q^2,\mu^2;t) = f_g(x,x',Q^2,\mu^2)F_N(t), \label{eq:rat5a} \ee
where we parameterize the form factor of the proton vertex by the form $F_N(t) = \exp(b_0 t)$
with $b_0 =2~\GeV^{-2}$ \cite{KMRsoft}.
In the domain specified by (\ref{eq:lmf_ineq}) the skewed unintegrated densities are given in terms of the
conventional (integrated) densities $g(x,Q_i^2)$. To single log accuracy, we have~\cite{MR01}\footnote
{In the actual computations we use a more precise form as given by eq.(26) of Ref.~\cite{MR01}.}.
\be f_g(x,x',Q_i^2,\mu^2) = R_g\frac{\partial}{\partial \ln Q_i^2} \left(\sqrt{T(Q_i,\mu)}\: xg(x,Q_i^2)\right),
\label{eq:rat6} \ee
where $T$ is the usual Sudakov form factor which ensures that the active gluon does not
emit additional real partons in the course of the evolution up to
the scale, $\mu$, of the hard process, so that the rapidity gaps survive.
This Sudakov factor $T$ is the result of resumming the
virtual contributions in the DGLAP evolution. It is given by
\begin{equation}
T(Q_\perp,\mu) = \exp\left(-\int_{Q_\perp^2}^{\mu^2}\frac{\alpha_S(k_t^2)}{2\pi}\frac{dk_t^2}{k_t^2}
\int_0^{1-\Delta}\left[zP_{gg}(z) + \sum_q P_{qg}(z)\right]dz\right). \label{eq:T}
\end{equation}
Here, in analogy to \cite{KKMRext}, we go beyond the collinear approximation 
and in the $T$ factor we resum, not just the single
collinear logarithms, but the single soft $\log\, 1/(1-z)$ terms as well. To a good approximation this can
be achieved by taking the upper limit of the $z$ integration in (\ref{eq:T}) to be
\be \Delta = \frac{k_t}{k_t + 0.62M_{\chi}}\,. \label{eq:monday1} \ee
The square root in (\ref{eq:rat6}) arises because the bremsstrahlung survival probability $T$ is only relevant to
 hard gluons. $R_g$ is the ratio of the skewed $x'\ll x$ integrated distribution to the conventional diagonal
density $g(x,Q^2)$. For $x\ll 1$ it is completely determined~\cite{SGMR}.

To compare expressions (\ref{eq:PPfusion}) and (the square of) (\ref{eq:rat3}),
we recall, assuming Regge factorization, that the unintegrated gluon density can
be written in the form
\be f_g(x,x',Q^2,\mu^2;t) = R_g~x^{\lambda(Q^2,t)}~\phi(Q^2,\mu^2)~g_N(t), \label{eq:fgR} \ee
These factors $x_i^\lambda$, in the $f_g$'s in (\ref{eq:rat3}), play the
role of the Regge factors $x_i^{1-\alpha_P(t_i)}$ in (\ref{eq:PPfusion}).
Moreover the form factor $F_N(t)$ in (\ref{eq:rat5a}) describes the $t$ dependence of the coupling
$g_N(t)$, while the remainder of the integral in (\ref{eq:rat3}) gives the
product $Vg_N^2(0)$.

\section{The cross section for exclusive diffractive $\chi$ production}

The uncertainties in the perturbative QCD predictions for the exclusive double-diffractive production
of heavy (Higgs) states of mass 120 GeV were discussed in Ref.~\cite{KKMRext}.
The uncertainties come from the infrared contribution to the $Q_\perp$ integral, (\ref{eq:rat3}),
from the choice of factorization scale, from the lack
of precise knowledge of the gluon at low scales and small $x$, from
deviations from formula (\ref{eq:rat6}) for large-angle gluon emission and
from the uncertainty in the calculation of the screening correction, $\hat{S}^2$.
For a Higgs $0^+$ state the total uncertainty was estimated to be a factor of
almost 2.5, that is up to almost 2.5, and down to almost 1/2.5 the quoted
value.  In Ref.~\cite{KKMRext} it was also estimated that the uncertainties for producing a heavy
$0^-$ state would be larger by almost another factor of 2.5.

For the exclusive double-diffractive production of light $\chi$ states the infrared uncertainties
will be even larger.
We computed the integral in (\ref{eq:rat3}) using the GRV94 parametrization \cite{GRV94}--
the only available parton set which extends down to a very low scale ($Q^2 \simeq 0.4{\rm ~GeV}^2)$.
To be safe, we choose a rather low energy, $\sqrt s$ = 60 GeV, which corresponds to $x\sim 0.05$,
where the GRV, MRST, CTEQ gluons are approximately equal; also where the $pp$ total
cross section is still approximately flat, which provides a
simple normalization of the Pomeron-nucleon vertex.  We find that half of the contribution to
the amplitude (\ref{eq:rat3}) comes from the region $Q_\perp < 0.85$ GeV.  Clearly perturbative
calculations are not justified in this domain. For low $Q_\perp$ the gluon propagator will be
modified by non-perturbative dynamics; for example by the presence of $G^a_{\mu\nu}G^a_{\mu\nu}$
condensates, effective gluon masses, confinement forces etc.  It appears that these dynamics effectively
suppress the low $Q_\perp$ contribution; for recent reviews see, for example, Ref.~\cite{Alk}.
Therefore the predictions below should be regarded as just an
order-of-magnitude indication of the expected rates.
If we evaluate amplitude (\ref{eq:rat3}) using GRV94HO partons \cite{GRV94} then we find, for $\sqrt s$ = 60 GeV,
that the `bare' cross section is
\begin{equation}
\left. \frac{d\sigma}{dp^2_{1\perp}dp^2_{2\perp}dy} \right|_{p_{1\perp}=p_{2\perp}=0,~y=0}~~\simeq~~8~\mu {\rm b/GeV}^4,
\label{eq:numT}
\end{equation}
whereas if the contribution from the region $Q_\perp < 0.85$ GeV is neglected then
we obtain 2 $\mu{\rm b/GeV}^4$.  By the bare cross section we mean the prediction before
including the rapidity gap survival factor ${\hat S}^2$.

We emphasize that the diagram of Fig.~1(a) plays the dominant role
when the transverse size of the Pomeron (that is the transverse
separation of the $x$ and $x'$ gluons) is much larger than the
size of the produced boson.  This is the case for Higgs
production\footnote{Indeed it has been shown \cite{LM} that there
is a negligible contribution from graphs where both gluons
(forming the Pomeron) couple to the top-quark loop.}, but it is
not so clear for $\chi_c$ production.  The other extreme is to
assume that the transverse size of the Pomeron is relatively
small, so that the Pomeron couples to each individual quark.  This
extreme corresponds to the `additive quark model'.  In this case
the fusion vertex may be calculated from the Feyman diagram shown
in Fig.~3.   The coupling of the charm quarks to form the $\chi_c$
meson is described in terms of the $\chi_c$ meson wave function,
$\psi(r)$, for which we apply the standard non-relativistic
formalism, see, for example, Refs.~\cite{BGK,Kuhn}.

\begin{figure}
\begin{center}
\centerline{\epsfxsize=7cm\epsfbox{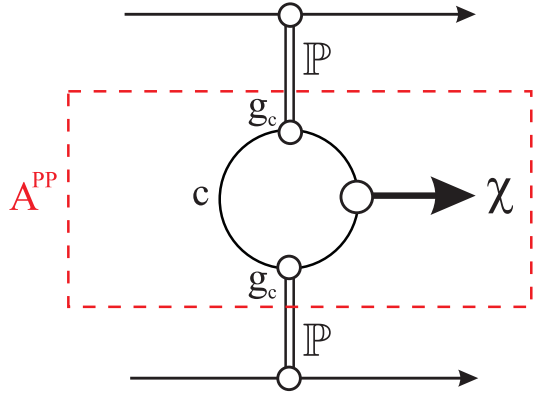}} \caption{The diagram used to
calculate the non-perturbative contribution 
to the exclusive process $pp \to p+\chi+p$. The amplitude, $A^{PP}$, describing the
Pomeron-Pomeron $\to \chi$ subprocess, is controlled by
the charm quark loop.  It is evaluated using (\ref{eq:APP2}). \label{fig:3a}}
\end{center}
\end{figure}

Next we have to estimate the coupling $g_c$ of the charm quark to the Pomeron.
For this, we can use the analysis of the data for charmonia production on nuclei \cite{cdata}.
The analysis \cite{Kop} showed that the cross section for $J/\psi$ on protons
can be parametrised as\footnote{The values of the cross section are in
qualitative agreement with the analysis of Ref.~\cite{FGSZ}.}
\begin{equation}
\sigma~=~\sigma_0~(s/s_0)^\Delta,
\label{eq:sigcp}
\end{equation}
with $\Delta=0.21$ and $\sigma_0=3.6$ mb, for $s_0 =100~{\rm
GeV}^2$. Thus the additive quark model estimate of the charm
quark-proton cross section\footnote{Note, however, a similar
analysis of $\psi ^{'}$ gives a somewhat larger value of $\sigma
(cp)$.} is $\sigma (cp) \simeq 2$ mb, for $s_{\psi p} \simeq
200~{\rm GeV}^2$, which corresponds to $\sqrt s =$~60 GeV of
(\ref{eq:numT}).  Phenomenologically, the Pomeron-quark coupling
may have a scalar or Dirac $\gamma _{\mu}$ matrix form.   The
latter means that Pomeron exchange is similar to photon exchange.
Thus it is reasonable to estimate the $PP-\chi$ coupling using the
known $\chi \to \gamma\gamma$ decay width, $\Gamma(\chi (0^+) \to
\gamma\gamma)\simeq 3$~keV \cite{RPP}.  However the analogy between
the Pomeron and the photon cannot be exact.  First, the Pomeron
has even $C$-parity.  Moreover the Pomeron is not a gauge field
and so its $\gamma _{\mu}$ vertex would not correspond to a
conserved current \cite{close}.   Indeed, unlike $\chi \to
\gamma\gamma$, for $PP \to \chi$ there are no graphs with a
permutation of the two Pomeron vertices.  The reason is that in
Regge theory all interactions are strongly ordered in rapidity
space.  Thus we omit the second term in eq.(2) of
Ref.~\cite{Kuhn}.

If we follow the above procedure, then, for $\chi_c (0^+)$,
we find that the $PP \to \chi$ vertex, $V$ of (\ref{eq:PPfusion}), is
\begin{equation}
V~\equiv~A^{PP}~=~g_c^2~\frac{a}{\sqrt 6}~\frac{4}{m_c},
\label{eq:APP}
\end{equation}
whereas the $\chi \to \gamma\gamma$ decay amplitude is
\begin{equation}
A^{\gamma \gamma}~=~\frac{4}{9}~(4\pi \alpha_{QED})~\frac{a}{\sqrt 6}~\frac{12}{m_c},
\label{eq:Agg}
\end{equation}
where $g_c$ is the Pomeron-c quark coupling, and $a=\sqrt {3/4\pi m_c}~\psi^{'}(0)$, where
$\psi^{'}(0)$ is the value of the first derivative of the bound-state wave function at the origin.
In this way we can normalize the $PP \to \chi$ vertex in terms of the known
$\chi \to \gamma\gamma$ decay width.  We obtain
\begin{equation}
A^{PP}~=~\frac{3g_c^2}{16\pi \alpha_{QED}}~A^{\gamma \gamma}.
\label{eq:APP2}
\end{equation}
If we were to choose a scalar Pomeron-quark coupling, then we would obtain an $A_{PP}$
amplitude which is a factor 5/4 larger.  However we prefer the $\gamma _{\mu}$ form of the vertex,
which conserves the $s$-channel helicity of the quark.

If, now, we take $\sigma(cp) = 2$~mb, and normalize the $A^{PP}$
amplitude as in (\ref{eq:APP2}),  then we find that
the  non-perturbative contribution\footnote{We call it ``non-perturbative" although, strictly speaking,
the true underlying dynamics of the $\sigma(cp)$ interaction is not well established.}
 to exclusive double-diffractive $\chi_c$ cross section is
\begin{equation}
\left.\frac{d\sigma^{\rm nonpert}}{dp^2_{1\perp}dp^2_{2\perp}dy}
\right|_{p_{1\perp}=p_{2\perp}=0,~y=0}~\simeq~1.1~\mu {\rm b/GeV}^4  \times {\hat S}^2,
\label{eq:cchi}
\end{equation}
with $t_i \simeq -p^2_{i\perp}$.
Note that the contribution of Fig.~3 is real (if
we assume Re/Im$\ll$1), and therefore does not interfere with the
imaginary amplitude of Fig.1.  Thus we simply add these two contributions
to the cross section. To be conservative, we only include the
perturbative contribution of gluons with  $Q_\perp > 0.85$ GeV: that is
\begin{equation}
\left.\frac{d\sigma^{\rm pert}}{dp^2_{1\perp}dp^2_{2\perp}dy}\right|_{p_{1\perp}=p_{2\perp}=0,~y=0}~~
\simeq~~2~\mu {\rm b/GeV}^4\times {\hat S}^2,
\label{eq:pert}
\end{equation}
at $\sqrt s$ = 60 GeV; see the discussion below (\ref{eq:numT}).

Now we need to consider the energy dependence.  First we note that the data for
$J/\psi$ photoproduction \cite{Jdata} are described by a Pomeron with
$\alpha_P (0)-1 \simeq 0.2$.  The $c{\bar c}$ system in this process has a similar mass
to the subprocess we are concerned with and so we take $\alpha_P (0)-1=\Delta = 0.2$.
The same value of $\Delta$ describes the behaviour of the $J/\psi$-proton cross
section, and is also consistent with the small $x$ behaviour found in the global
parton analyses of deep inelastic scattering data.
Adding together the above perturbative and non-perturbative contributions at
$\sqrt s$ = 60 GeV, and using the appropriate energy dependence, $s^{2\Delta}$ of (\ref{eq:sdep}), 
we find the cross section for
$pp \to p+\chi_c+p$, at the Tevatron energy, ${\sqrt s} =  2$~TeV, is
\begin{equation}
\left. \frac {d\sigma(\chi_c)}{dy}\right|_{y=0}~=~130~{\rm nb}.
\label{eq:total}
\end{equation}
Here we have included the survival factor ${\hat S}^2=0.07$ of the rapidity gaps, (\ref{eq:S2aTev}),
and integrated over the transverse momenta $p_{i\perp}$.
To obtain an approximate expression for the $t$ dependence of the exclusive double-diffractive cross section,
we neglect the possible difference in the form factors corresponding to Figs.~1 and 3, 
and the effects of the screening corrections on the $p_{i\perp}$ distributions, which results in
the behaviour
$F_N(t_1)^2 F_N(t_2)^2 ~{\rm exp}(\alpha^{'}(t_1+t_2){\rm ln}(s/M^2_{\chi})).$

To obtain the total observable cross section at the Tevatron we integrate over an effective
rapidity interval of $\Delta y=5$ which gives
\begin{equation}
\sigma_{\rm excl}(\chi_c)~=~650~{\rm nb}.
\label{eq:totalint}
\end{equation}
This, more detailed estimate, is very close to our previous result \cite{KMRmm}.  

Note that
the production of a (non-relativistic) $2^{++}$ quarkonium state
may allow the discrimination between the two mechanisms
shown in Fig.~1 and Fig.~3.
As we already discussed (see Refs.~\cite{KMRmm,FY} for more details)
in the on-shell approximation the $J_z(gg)=0$ amplitude vanishes.
Then, the only contribution comes
from the amplitude for $|J_z|=2$ quarkonium ($2^{++})$ production, arising from
the violation of the $J_z=0$ selection rule due to non-zero $p_\perp$
of the protons. Thus, we expect that the amplitude corresponding
to Fig.~1 vanishes for forward $2^{++}$ quarkonium production.
On the other hand there is no selection rule similar to
$J_z(gg)=0$ for the general amplitude of Fig.~2 (though even P and C
still hold). In particular, the amplitude of Fig.~3 allows $2^{++}$ production
at any $p_\perp$ and, therefore, does not vanish as $p_\perp \to 0$.
Another difference between the contributions of Fig.~1 and Fig.~3,
is that, in Fig.~3, a centrally-produced
$\chi(2^{++})$ meson has $J_z=0$, while for Fig.~1 
non-relativistic $2^{++}$ quarkonium has $|J_z|=2$.

For
completeness we gather together all our predictions for the cross sections for
double-diffractive $\chi$ production in Table 1.  The evaluation of the inclusive
cross section is discussed in the next Section, and also in Section 8.  Comments
on $\chi_b$ may be found in Section 6. Although the cross sections are large, we note that
the relevant $\chi$
decay branching fractions are small.  Traditionally, the experimentally favoured 
decay is $\chi_c \to \gamma J/\psi \to \gamma\mu^+\mu^-$ with branching fraction $6 \times 10^{-4}$.
On the other hand,  $\chi_c \to (\pi^+\pi^-+K^+K^-)$ has branching fraction 0.011, and this 
2--prong configuration could prove more favourable.   
For $\chi_b$ the corresponding branching
ratios are not so well known, but it hard to imagine that the situation will be better.
Recall that $\chi_b$ decays dominantly into two gluon jets.
Unfortunately this decay channel will suffer from a large  background from 
QCD dijet production in the
same kinematical configuration.
Using the results of Ref.~\cite{KMRProsp}, and
neglecting the K-factor in the background
calculation, we estimate the signal-to-background ratio
 to be  $S/B \sim  5 \times 10^{-3} /\Delta M ({\rm GeV})$.
Certainly we would expect  that there is
a sizeable (on the $\sim 1\%$ level) branching ratio
for  $\chi_{b}(0^+) \to \gamma + \Upsilon$ decay. But at the moment there is no
experimental data on this channel. Therefore the situation regarding detection
of the $\chi_b(0^+)$ in the central detector remains unclear.

What are the uncertainties in our cross-section predictions?  First, it is interesting to note that the results
are almost independent on the choice of the slope, $b=2b_0+2\alpha^{'}{\rm ln}1/x$,
of the $p_{i\perp}$ distributions.  The reduction caused by a larger slope is largely
compensated by an increased survival factor ${\hat S}^2$ for the more peripheral reaction.
For example, varying the slope $2b_0$ from 4 to 6 ${\rm GeV}^{-2}$ changes the individual
values of ${\hat S}^2$ and $b^2$ by approximately a factor of 2, but decreases ${\hat S}^2/b^2$
by only about 10-15\% at the Tevatron and the LHC energies\footnote{For the same reason
the cross section predictions for exclusive double-diffractive Higgs production are practically
independent of the value taken for the slope $b$.}.
On the other hand, the cross section would be 4 times smaller if we were to take a `soft' Pomeron
intercept,  $\alpha_P (0)-1=\Delta = 0.1$, instead of 0.2, for the extrapolation from
$\sqrt s$ = 60 GeV to the Tevatron energy.  Moreover the charm quark cross section may be
larger (see the footnote below (\ref{eq:sigcp})), and there may be an additional
contribution from the low $Q_\perp$ domain, $Q_\perp < 0.85$ GeV, of the gluon-induced
amplitude (\ref{eq:rat3}).  Thus the cross section estimate of (\ref{eq:totalint})
could be a factor of 4 higher.

\begin{table}$$\begin{array}{|c|cc|cc|} \hline
 \rule[0ex]{0ex}{3ex}  & \multicolumn{2}{c|}{{\rm Tevatron}\ \sqrt s =
2~{\rm TeV}} & \multicolumn{2}{c|}{{\rm LHC}\ \sqrt s = 14~{\rm
 TeV}}    \\
 & \rule[0ex]{4ex}{0ex} \chi_c \rule[0ex]{4ex}{0ex} & \rule[0ex]{4ex}{0ex}
\chi_b \rule[0ex]{4ex}{0ex} &
 \rule[0ex]{3.3ex}{0ex} \chi_c \rule[0ex]{3.3ex}{0ex} &
\rule[0ex]{3.3ex}{0ex} \chi_b \rule[0ex]{3.3ex}{0ex} \\[1ex] \hline
 \rule[0ex]{0ex}{3ex} d\sigma_{\rm excl}/dy|_{y=0} & 130 & 0.2 & 340 & 0.6
\\
 \sigma_{\rm excl} & 650 & 0.5 & 3000 & 4 \\[1ex] \hline
 \rule[0ex]{0ex}{3ex} d\sigma_{\rm incl}/dy|_{y=0} & 13 & 0.06 & 30 & 0.2 \\
 \sigma_{\rm incl} & 70 & 0.3 & 200 & 2 \\[1ex] \hline
\end{array}$$
\caption{The cross sections (in nb) for exclusive and inclusive double-diffractive $\chi_c$ and $\chi_b$
production at the Tevatron and LHC.  From the discussion in the text, we
see that these predictions should be regarded as order of magnitude
estimates.  The accuracy is limited because of the low scale associated
with $\chi$ production.}
\end{table}

\section{The rates for inclusive double-diffractive $\chi$ production}

The cross section for double-diffractive inclusive production is
calculated is an analogous way to the exclusive one. Indeed, it may be
considered as `exclusive' $\chi$ meson production in
parton-parton collisions. The flux of initial partons
is given by the known parton distributions inside a proton, while
the unintegrated skewed distributions (inside an
individual) parton, $f_g$, are calculated in terms of the non-forward
BFKL amplitudes. It is important to note that here there is no form factor for
point-like parton scattering and that there are sizable contributions over
a much larger range of transverse
momenta $p_{i\perp}$, with $i=1,2$. Therefore we keep just the leading-order
contributions; that is, the integrals over $Q_\perp,~p_{1\perp}~{\rm and}~p_{2\perp}$ take
logarithmic forms. In particular
\begin{equation}
\frac{d^2Q_\perp\ P(\chi (0^+))}{Q^2_\perp (\vec Q_\perp - \vec p_{1\perp})^2(\vec Q_\perp + \vec
p_{2\perp})^2}~~{\rm takes~the~form}~~  \frac{d^2Q_\perp}{Q^2_\perp}~
\frac{\vec p_{1\perp}\cdot\vec p_{2\perp}}{ p_{1\perp}^2~ p_{2\perp}^2},
\label{eq:logamp}
\end{equation}
so that the final integrals in the expression for the cross section become
\begin{equation}
 \frac{d^2Q_\perp}{Q^2_\perp}~\frac{d^2Q_\perp^{'}}{Q_\perp^{'2}}~\frac{d^2p_{1\perp}}{p^2_{1\perp}}~\frac{d^2p_{2\perp}}{p^2_{2\perp}};
\label{eq:logint}
\end{equation}
see Refs.~\cite{KMRdijet,KMRProsp} for details.

As mentioned before, for a heavy centrally produced state, such as a Higgs boson, the
cross section for inclusive double diffractive production is predicted to be much larger than that for
exclusive production.  The reasons are as follows:

(a) due to the absence of proton form factors in the inclusive process there are sizeable
contributions to the $dp_\perp^2/p_\perp^2$ integral with $Q_\perp<p_\perp<M_H/2$,

(b) the $T$-factor suppression is weaker for inclusive production, as now $T(p_\perp,M_H)>T(Q_\perp,M_H)$,

(c) for inclusive production there is no P-even $J_z=0$ selection rule \cite{Liverpool,KMRItal,
KMRmm,KMRProsp}, which suppresses\footnote{The exclusive double-diffractive production
of vector and axial vector states is suppressed as a consequence of the
Landau-Yang theorem \cite{LY} for massless gluons.}, for example,
$0^-, 2^+$ (non-relativistic) quarkonium exclusive production \cite{KMRmm,FY,KMRtagg}.

On the other hand for the relatively light $\chi$ states the typical values
of $p_\perp$ transferred through the Pomeron, and $Q_\perp$ of the internal gluon (inside
the Pomeron) are of the same order.  As a consequence inclusive production is not strongly enhanced
in comparison with exclusive production.
Insight into this result can be obtained from the discussion leading to (\ref{eq:inexcl}).
It is therefore not surprising that, using the perturbative QCD approach of the previous section,
we estimate that
\be
\sigma_{\rm incl}(\chi_c(0^+)) \simeq 70~{\rm nb}
\label{eq:A5}
\ee
at the Tevatron energy $\sqrt s$~=~2 TeV, which is about 10 times less than the exclusive
double-diffractive cross section $\sigma_{\rm excl} \simeq 650$~nb.  It is
interesting to note that the perturbative QCD estimate reflects the smallness
of the triple-Pomeron vertex.  The low QCD estimate arises from the small gluon
density (at low $x$ and small scale) obtained in the global parton analyses of
deep-inelastic and related data.
As we have already mentioned, recently the CDF collaboration \cite{CDFchi} 
have reported their first results on
 $J/\psi + \gamma$ central production,
 with rapidity gaps on either side, and with a mass consistent
 with that of $\chi_c$. These results are still
preliminary, but the observed cross section
 is compatible with  the expectations of Ref.~\cite{KMRmm}, assuming that
the background issues are well understood and turn out to be small.
 With more data to come, and with a study of the background, it will be possible to
 obtain more detailed information on this process.

In going from the Tevatron to the LHC energy, $\sqrt s$~=~14 TeV, the double-diffractive
inclusive cross section (for a fixed rapidity gap interval), 
$\sigma_{\rm incl}$, is expected to grow as $s^{\Delta}$, with
$\Delta ~\simeq~0.2$.  Now we have to integrate over the rapidity gap
intervals $\Delta\eta > 3$, that is over the masses $dM^2_1/M^2_1$
and $dM^2_2/M^2_2$.
In the  limit of  small fixed masses  $M_1$, $M_2$ the
integrand grows faster -- as $s^{2\Delta}$ -- similar to  the elastic
cross section.
On the other hand the rapidity gap survival factor $\hat{S}^2$
decreases as $s^{-0.1}$ \cite{KMRsoft}, see (\ref{eq:S2sdep}). To be explicit, we find \cite{KMRsoft}
$$\hat{S}^2({\rm LHC})/ \hat{S}^2({\rm Tevatron})~\simeq~2/3.$$
We see that the energy dependence of the cross section and the behaviour of the survival factor
partially compensate each other, so
\be
\frac {d\sigma_{\rm incl}}{dy}({\rm LHC})~\simeq~2.5~\frac {d\sigma_{\rm incl}}{dy}({\rm Tevatron}),
\label{eq:A6}
\ee
for $y=0$. At the LHC, a larger rapidity interval is available, $\Delta y \sim 8$, which leads to
an increased cross section
\be
\sigma_{\rm incl}(\chi_c(0^+)) \simeq 200~{\rm nb}.
\label{eq:A6a}
\ee
For the exclusive process, the cross section $d\sigma /dy$ is
proportional to $s^{2\Delta}$. In fact the QCD
estimate gives $\sigma_{\rm excl}(\chi_c(0^+)) \simeq 3000~{\rm nb}$
at the LHC.   These predictions are collected together in Table 1.

\section{Predictions for double-diffractive $\chi_b$ production}

The perturbative QCD predictions of the cross sections of double-diffractive $\chi_b$
production are more justified than those for $\chi_c$, on account of the larger $\chi_b$ mass.
For this reason we only evaluate the perturbative amplitude (\ref{eq:rat3}), which
corresponds to the diagrams of Fig.1.  Moreover,
as before, we keep just the contribution from the $Q_\perp > 0.85$ GeV integration region.
This is found to give about $\frac{2}{3}$ of the total amplitude (\ref{eq:rat3}) obtained using GRV94HO
partons \cite{GRV94}. The energy dependence of the bare cross section is driven by the small
$x$ behaviour of the gluon.  However we use again $\Delta=0.2$, which is much more consistent
with the recent partons sets (for example \cite{MRST99}) than the GRV94 set. 

Another uncertainty is the absence of an experimental measurement of the gluonic width,
$\Gamma(\chi_b(0^+) \to gg)$.   Potential models \cite{chibV} predict a bare width of
$\Gamma_0(\chi_b \to gg) = 1.3-1.9$ MeV, which should be multiplied by the NLO
correction factor $K=(1+9.8\alpha_S/\pi)$ \cite{K}.  This is some 5 times larger
than the prediction $\Gamma_0(\chi_b \to gg) \simeq 350$ keV obtained from
QCD lattice calculations \cite{lattice}.   For the predictions of the cross sections of
double-diffractive
$\chi_b$ production at the Tevatron and LHC, which we show in Table 1,
we have taken $\Gamma(\chi_b \to gg)$ = 1.3$K$ MeV.  Note that these predictions
are larger than those given in Ref.~\cite{KMRmm}, where the lattice estimate of
the width was used\footnote{Note, however, that the calculations of the matrix elements
in Ref.~\cite{lattice} were based on the lattice NRQCD results of
Ref.~\cite{BSK}, which used quenched approximation.
As was shown in \cite{BSK1}, this approximation underestimates the
NRQCD matrix elements. Thus the values of the widths, in the case of three
light quark flavours, should be higher than the quenched results.}; see also Ref.~\cite{FY}.

\section{Implications of the larger exclusive rate for $\chi_c$ production}

The QCD predictions of the double-diffractive exclusive and inclusive $\chi_c$ production show
that the former process dominates.  Indeed at the LHC we estimate
\be
\sigma_{\rm excl}/\sigma_{\rm incl}~\sim~10.
\label{eq:A8}
\ee
It means that by just selecting events with two rapidity gaps we will observe mainly
the exclusive process, even without requiring the forward protons to be tagged!

One topical application would be to search for the exclusive production of
the new $X(3872)$ charmonium state observed recently by the Belle and CDFII collaborations \cite{chi3872}.
If this state were a radial excitation of the $\chi_c(0^+)$ meson, then it would
be observed with a cross section
\be
\sigma (pp \to p+X+p)~\sim~100 (500)~{\rm nb}
\label{eq:A9}
\ee
at
the Tevatron (LHC).  Of course, looking for a specific decay
$X \to \psi \pi^+ \pi^- \to \mu^+ \mu^- \pi^+ \pi^-$ will reduce the signal
by about 1000, but still the cross section is large.

 Since the $X$ production rate is proportional to the $X \to gg$ decay width,
a measurement of the $pp \to p+X+p$ cross section can give an estimate of the
$X \to gg$ coupling.  To reduce the (large) theoretical uncertainties it
would be better to compare the rates for $X(3872)$ and $\chi_c(3415)$, as
the $gg$ width is known for the latter state\footnote{The effective $gg^{PP}$
luminosity, ${\cal L}(M^2)$, is essentially flat in this mass interval.}.

If $X(3872)$ were a $D\bar{D^*}$-molecule (see, for instance, \cite{DD}), then
the exclusive signal would probably not be seen\footnote{In this case there
should be some contribution from the $D$-meson loop, analogous to the
charm quark loop in Fig.~3, but it is expected to be small.}.
Also note that, due to the P-even,
$J_z=0$ selection rule\footnote{Unfortunately the $J_z=0$ selection rule
is less precise at these low scales.  The admixture of $|J_z|=2$ states may be
up to 20-30\%. However, it is encouraging that expectations based on this rule
appear to be in good qualitative agreement with the available data on double-diffractive
meson production; see \cite{KMRtagg} for details.}, the exclusive cross section is suppressed for
the double-diffractive production of $0^-,~2^+$ or C-odd
bosons.  Thus by observing the $X(3872)$ state in the exclusive process we
would conclude that it is a C-even, and most probably a $0^+$, particle.

\section{Parity determination}

With sufficient luminosity we can use the inclusive process, (\ref{eq:A2}),
to study $0^-$ or $2^{++}$ production by Pomeron-Pomeron fusion.  If the
transverse momentum $p_\perp$ transferred through the Pomeron is small,
$p_\perp \ll Q_\perp$, then the inclusive process is controlled by the
same P-even, $J_z=0$ selection rule as exclusive production.  However if
we select events with large $p_{i\perp}$ then the production of
P-odd and $J_z=\pm 2$ states become possible.  Note that large
$p_\perp$'s can be measured as the transverse energy flows,
$p_{1\perp}=-E_{1\perp}$ and $p_{2\perp}=E_{2\perp}$, in the proton fragmentation regions.
Moreover the major part of the energy flow is carried by one jet (with
lowest rapidity $|y|$ in the centre-of-mass frame).  Indeed at lowest
order this jet carries the entire $E_{i\perp}$.  These measurements allow
a study of the $\phi$ dependence, where $\phi$ is the azimuthal angle
between the energy flows $\vec{E}_{1\perp}$ and $\vec{E}_{2\perp}$.
The $\phi$ distribution depends on the parity of the
produced system \cite{KKMRext}:
\be
d\sigma/d\phi~\propto ~1~+~{\rm cos} 2\phi~~~~~~~~{\rm for~a~natural~parity~state,}~P=(-1)^J
\label{eq:A10}
\ee
whereas
\be
d\sigma/d\phi~\propto ~1~-~{\rm cos} 2\phi~~~~~~~~{\rm for~an~unnatural~parity~state,}~P=-(-1)^J.
\label{eq:A11}
\ee
We emphasize, that since we consider rather large $E_\perp$, these angular
distributions are less sensitive to soft rescattering.

To get some idea of the event rates, we present the cross sections for
$pp \to X + \chi(0^+) + Y$ in Table 2, for different choices of the $E_\perp$ cut.
If we take $E_{i\perp}>3$ GeV at the Tevatron, then $\sigma_{\rm incl}(\chi_c(0^+))\sim 0.5$ nb,
whereas for $E_{i\perp}>7$ GeV at the LHC we estimate $\sigma_{\rm incl}(\chi_c(0^+))\sim 40$ pb.
Note that the reliability of the predictions with large $E_\perp$ cuts is better, since we do not
enter the infrared domain.    These are rather large cross sections, but recall that the
relevant decay branching fractions are small, see Section~4.

\begin{table}$$\begin{array}{|l @{\qquad} c c c|} \hline
 & E_\perp>3 & E_\perp>5 & E_\perp>7  \rule[-1.5ex]{0ex}{5ex}  \\ \hline
 {\rm Tevatron} &  &  & \\
 \sigma(\chi_c) & 500 & 20 & \\
 \sigma(\chi_b) & 35  & 5 & \\ \hline
 {\rm LHC} & & & \\
 \sigma(\chi_c) & 6000 & 400 & 40 \\
 \sigma(\chi_b) & 500 & 100 & 30 \\ \hline
\end{array}$$
\caption{The inclusive double-diffractive cross sections (in pb) for various choices
of the cut on the transverse energy flows, $E_{i\perp}$ (in GeV).
To account for the large gluon virtuality, which occurs for large $E_{i\perp}$, we
follow a procedure similar to that used in Ref.~\cite{Kuhn} for the calculation
of the amplitude for the decay of $\chi(0^+)$ into virtual photons.}
\end{table}

Finally, note that the ratio of the $0^{++}$ and $2^{++}$ production cross sections is
\be
\frac{\sigma_{\rm incl}(2^{++})}{\sigma_{\rm incl}(0^{++})}~=~\frac{5}{4}~
\frac{\Gamma(2^{++})}{\Gamma(0^{++})},
\label{eq:A12}
\ee
where we have neglected the mass difference between the $M(2^{++})$ and $M(0^{++})$ mesons.

\section{Conclusions}

We find that both the Regge formalism and perturbative QCD predict essentially the
same qualitative behaviour for the central double-diffractive production of
`heavy' $\chi_c(0^{++})$ and $\chi_b(0^{++})$ mesons. Due to the
low scale, $M_\chi /2$, there is a relatively small contribution coming from the process 
in which the incoming protons dissociate. Therefore simply 
selecting events with a rapidity gap on either side of the $\chi$, almost
ensures that they will come from the exclusive reaction, $pp\to p\ +\ \chi\ +\ p$.

 We evaluated the expected double-diffractive cross sections and demonstrated that
they are sufficiently large for both $\chi_c$ and $\chi_b$ meson production to
be observed.  Since the rapidity-gap survival factor, $\hat{S}^2$, decreases with energy, the
cross section $d\sigma/dy$ at the LHC energy only exceeds that at the Tevatron
energy by a factor of about 3.   Our cross section predictions should
be regarded as only order of magnitude estimates, because of the low scale,
nevertheless these processes can be very informative. For example, the observation of the new
charmonium state $X(3872)$ in the exclusive process $pp \to p+X+p$ would be a strong
argument in favour of its  quantum numbers being $J^{PC}=0^{++}$. Moreover,
since the exclusive cross section is proportional to the gluonic width,
it will be possible to measure the width $\Gamma(X\to gg)$ by comparing
the rates of exclusive $\chi_c$ and $X$ production. While there is a sizeable uncertainty
in the predictions for the overall {\it rates}
 of double difffractive $\chi$ production, we stress that the
kinematic distributions, for example the $\chi$ transverse momentum distributions,
should be more reliable.

Although exclusive $\chi$ production is expected to dominate, the event rates should be large
enough to select double-diffractive dissociative events with large transverse
energy flows in the proton fragmentation regions. Such events are particularly
interesting.  First, in this case, the large value of
$E_\perp$ provides the scale to justify the validity, and the reasonable
accuracy, of the perturbative QCD calculation of the cross section. Next, by measuring the
azimuthal distribution between the two $E_\perp$ flows we can determine the
parity of the centrally produced system.

An interesting extension of the exclusive double-diffractive
approach, would be to observe central open $\bb$ production; 
namely $b,{\bar b}$ jets with $p_\perp \gapproxeq m_b$.
Again, this would put the application of perturbative QCD on a sounder footing. 
It would allow a check of the perturbative formalism, as well as
a study of the dynamics of $\bb$ production.

\section*{Acknowledgements}

We thank Mike Albrow, Albert De Roeck, Risto Orava and Angela Wyatt for useful discussions.
ADM thanks the Leverhulme Trust for an Emeritus Fellowship and MGR thanks the IPPP at the University of
Durham for hospitality. This work was supported by
the UK Particle Physics and Astronomy Research Council, by grant RFBR 04-02-16073
and by the Federal Program of the Russian Ministry of Industry, Science and Technology
SS-1124.2003.2.

\end{document}